\documentclass{aa}

\usepackage{natbib,color}
\bibpunct{(}{)}{;}{a}{}{,} 

\usepackage{graphicx,url}
\usepackage{mathrsfs,amssymb,amsmath}

\usepackage[varg]{txfonts}

\newcommand{\ud}{\mathrm{d}}
\newcommand{\be}{\begin{equation}}
\newcommand{\ee}{\end{equation}}

\newcommand{\h}{{\rm H}}
\newcommand{\hh}{{\rm H_{2}}}
\newcommand{\HI}{\ion{H}{I}}
\newcommand{\hf}{\rm H_{\rm fast}}

\newcommand*\xbar[1]{%
   \hbox{%
     \vbox{%
       \hrule height 0.5pt 
       \kern0.2ex
       \hbox{%
         \kern-0.1em
         \ensuremath{#1}%
         \kern-0.0em
       }%
     }%
   }%
}

\begin{document}

\title{Production of atomic hydrogen by cosmic rays in dark clouds}

\author{Marco~Padovani\inst{1}, Daniele~Galli\inst{1}, Alexei~V.~Ivlev\inst{2}, Paola~Caselli\inst{2},
Andrea~Ferrara\inst{3}}

\authorrunning{M. Padovani et al.}


\institute{INAF--Osservatorio Astrofisico di Arcetri, Largo E. Fermi 5, 50125
Firenze, Italy\\
\email{[padovani,galli]@arcetri.astro.it}
\and
Max-Planck-Institut
f\"ur Extraterrestrische Physik, Giessenbachstr. 1, 85741 Garching, Germany\\
\email{[ivlev,caselli]@mpe.mpg.de}
\and
Scuola Normale Superiore, Piazza dei Cavalieri 7, 56126 Pisa, Italy\\
\email{andrea.ferrara@sns.it}
}


\abstract
{The presence of small amounts of atomic hydrogen, detected as absorption dips in the 21 cm line spectrum, is a well-known
characteristic of dark clouds. The abundance of hydrogen atoms measured in the densest regions of molecular
clouds can be only explained by the dissociation of $\hh$ due to cosmic rays.}
{We want to assess the role of Galactic cosmic rays in the formation of atomic hydrogen, by using
recent developments in the characterisation of the low-energy spectra of cosmic rays and advances in the modelling
of their propagation in molecular clouds.}
{We model the attenuation of the interstellar cosmic rays entering a cloud and compute the dissociation rate of
molecular hydrogen due to collisions with cosmic-ray protons and electrons as well as fast hydrogen atoms. We compare our
results with the available observations.}
{The cosmic-ray dissociation rate is entirely determined by secondary electrons produced in primary
ionisation collisions. These secondary particles constitute the only source of atomic hydrogen at column densities
above $\sim10^{21}$~cm$^{-2}$. We also find that the dissociation rate decreases with column density, while the
ratio between the dissociation and ionisation rates varies between about 0.6 and 0.7. From comparison with
observations we conclude that a relatively flat spectrum of interstellar cosmic-ray protons, as the one suggested by
the most recent Voyager~1 data, can only provide a lower bound for the observed atomic hydrogen fraction.
An enhanced spectrum of low-energy protons is needed to explain most of the observations.}
{Our findings show that a careful description of molecular hydrogen dissociation by cosmic rays can explain the
abundance of atomic hydrogen in dark clouds. An accurate characterisation of this process at high densities is
crucial for understanding the chemical evolution of star-forming regions.}


\maketitle

\section{Introduction}

The formation of molecular hydrogen occurs on dust grains in molecular clouds through the reaction between two
hydrogen atoms. Being an exothermic process, $\hh$ is then released into the gas phase. Depending on position in the
cloud (or the amount of visual extinction measured inward from the cloud's edge), two processes determine the destruction of
$\hh$ and the restoration of the atomic form: photodissociation due to interstellar (hereafter IS) UV photons and
dissociation due to cosmic rays (hereafter CRs). In the diffuse part of molecular clouds, UV photons regulate the abundance
of atomic hydrogen by dissociating $\hh$, while in the densest parts IS UV photons are blocked by dust absorption as
well as by $\hh$ line absorption \citep{hw71}. In the deepest parts of the cloud, CRs
dominate
the destruction of molecular hydrogen.

A wealth of studies have been carried out to characterise the origin of the atomic hydrogen component in dense environments
\citep[e.g.][]{ms78,wg88,mb95,lg03,gl05}, but the rate of CR dissociation was always assumed to be constant (i.e.,
independent of the position in the cloud) or simply neglected. In this paper we want to explore in more detail the role of
CRs -- especially after the latest data release of the Voyager~1 spacecraft \citep{cs16}, showing that the measured proton
and electron fluxes are not able to explain the values of the CR ionisation rate estimated in diffuse clouds
(e.g.~\citealt{in15}; \citealt{phan18}). In our previous work (e.g.~\citealt{pgg09}; \citealt{pg13}; \citealt{ph13};
\citealt{ip15}; \citealt{pi18}) we postulated the presence of a low-energy component in the IS CR proton spectrum, with
which it is possible to recover the high ionisation rates observed in diffuse clouds. 

We treat a cloud as a semi-infinite slab. Such a simplification is completely justified for our purposes, for the following
reasons. First, attenuation of IS UV photons occurs in a thin gas layer near the cloud's surface (with a visual
extinction of $A_{\rm V}\approx1-3$~mag), i.e., at column densities much smaller than those characterising the line-of-sight
thickness of a cloud. Second, CRs propagate through a cloud along the local magnetic field.
The latter assumption is always valid since 
the Larmor radius of sub-relativistic CRs is much smaller than any characteristic spatial scale of the cloud
\citep{pg11} and the correlation length of the magnetic field \citep{hh09}.
Therefore, irrespective of the field geometry, we can measure the coordinate along the local field line and treat the
problem as one-dimensional \citep{pi18}\footnote{The CR ionisation rate is then a function of the {\it effective}
column density, measured along the field line. To facilitate the presentation of our results, in this paper we assume the
line-of-sight and the effective column densities to be the same.}. One can straightforwardly generalise these considerations
to a slab of a finite thickness by adding IS particles entering the cloud from the opposite side; however, given a strong
attenuation, this addition is only important for clouds with
column densities of $\approx10^{22}$~cm$^{-2}$ or less 
(increasing the ionisation and dissociation rates in the cloud's centre by up to a factor of 2).

The paper is organised as follows:
in Sect.~\ref{h2diss} we discuss the main processes of $\hh$ dissociation by CR protons,
electrons, and fast hydrogen atoms, and carefully compute the resulting dissociation rate as a function of the column
density;
in Sect.~\ref{baleq} we present equations to compute the fractions of atomic and
molecular hydrogen;
in Sect.~\ref{comparison} we compare our theoretical findings with available observations; in
Sect.~\ref{discconc} we discuss implications for our outcomes and summarise the most important results.

\section{CR dissociation reactions with H$_2$}\label{h2diss}

\begin{figure}[!h]
\begin{center}
\resizebox{\hsize}{!}{\includegraphics{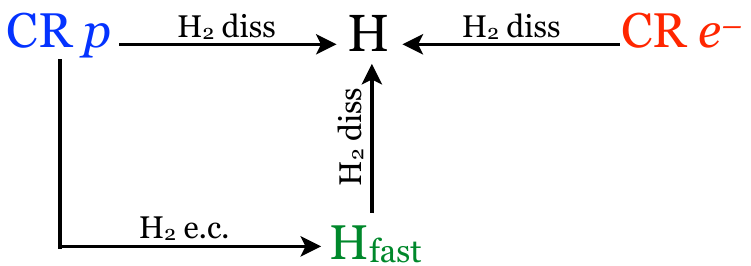}}
\caption{Dissociation diagram showing the three main processes of production of atomic hydrogen.
Labels ``diss'' and ``e.c.'' refer to dissociation and electron capture, respectively.}
\label{diagram}
\end{center}
\end{figure}

We consider dissociation processes induced by CR
primary and secondary electrons, CR protons, and fast hydrogen atoms colliding with
molecular hydrogen.
A schematic diagram of different dissociation paths is depicted in Fig.~\ref{diagram}.

\subsection{Electron impact}\label{disse}

Electrons can produce atomic hydrogen through excitation of five electronic states of the $\hh$ triplet
($a^{3}\Sigma_{g}^{+}$, $b^{3}\Sigma_{u}^{+}$, $c^{3}\Pi_{u}$, $e^{3}\Sigma_{u}^{+}$, and $d^{3}\Pi_{u}$)
followed by dissociation. 
While the radiative decay from the state $b^{3}\Sigma_{u}^{+}$ is fully dissociative, the decay from
$e^{3}\Sigma_{u}^{+}$ contributes to dissociation at 20\%, and dissociation from the other states is
negligible\footnote{There is also a contribution from the $\hh$ singlet state, but the respective cross section peaks at
about $40-50$~eV with a maximum value of $3.02\times10^{-18}$~cm$^{2}$, which is a factor $\approx20$ lower than the peak
value of the triplet-state cross section.}. Thus, the dissociation cross section by electron impact is given by
\begin{equation}
\sigma_{\rm diss}^{e}\simeq\sigma_{\rm exc}^{e}(X\rightarrow b^{3}\Sigma_{u}^{+})+ 0.2\sigma_{\rm exc}^{e}(X\rightarrow
e^{3}\Sigma_{u}^{+}).
\end{equation}

\subsection{Proton impact}\label{dissp}

Atomic hydrogen can also be produced by protons, by direct dissociation of H$_2$ from the vibrational state
$v=0$. The $\hh$ excitation cross sections by electrons, $\sigma_{\rm exc}^{e}$, and the dissociation cross section by
protons, $\sigma_{\rm diss}^{p}$, have been parameterised by \cite{janevbook} as
\begin{equation}\label{sdisspe}
\sigma(E) = \frac{a}{E^{\alpha_{1}}}\left[1-\left(\frac{E_{0}}{E}\right)^{\alpha_{2}}\right]^{\alpha_{3}}%
\times{\rm 10^{-16}~cm^{2}}\,,
\end{equation}
with the energy $E$ in eV. In Table~\ref{tab:coeffsigma} we list the values of factor $a$, exponents $\alpha_{1,2,3}$,
and the energy threshold $E_{0}$ for the respective cross sections.

\begin{table*}
\caption{Parameters for the proton dissociation cross section and the (relevant) electron excitation cross sections
(Eq.~\ref{sdisspe}).}
\begin{center}
\begin{tabular}{cccccc} \hline\hline
Reaction & $a$ & $\alpha_{1}$ & $\alpha_{2}$ & $\alpha_{3}$ & $E_{0}$~[eV]\\
\hline
$\!\!\!\!\!\!\!\!\!p+\hh\rightarrow p+\h+\h$ & $7.52\times10^{3}$ & 4.64 & 5.37 & 2.18 & 6.72\\
\hline
$e+\hh\rightarrow e+\hh^{*}(b^{3}\Sigma_{u}^{+})$ & $5.57\times10^{3}$ & 3.00 & 2.33 & 3.78 & 7.93\\
$e+\hh\rightarrow e+\hh^{*}(e^{3}\Sigma_{u}^{+})$ & $4.17\times10^{2}$ & 3.00 & 4.50 & 1.60 & 13.0\\
\hline
\end{tabular}
\end{center}
\label{tab:coeffsigma}
\end{table*}

\subsection{Fast hydrogen atom impact}

\begin{figure}[!h]
\begin{center}
\resizebox{\hsize}{!}{\includegraphics{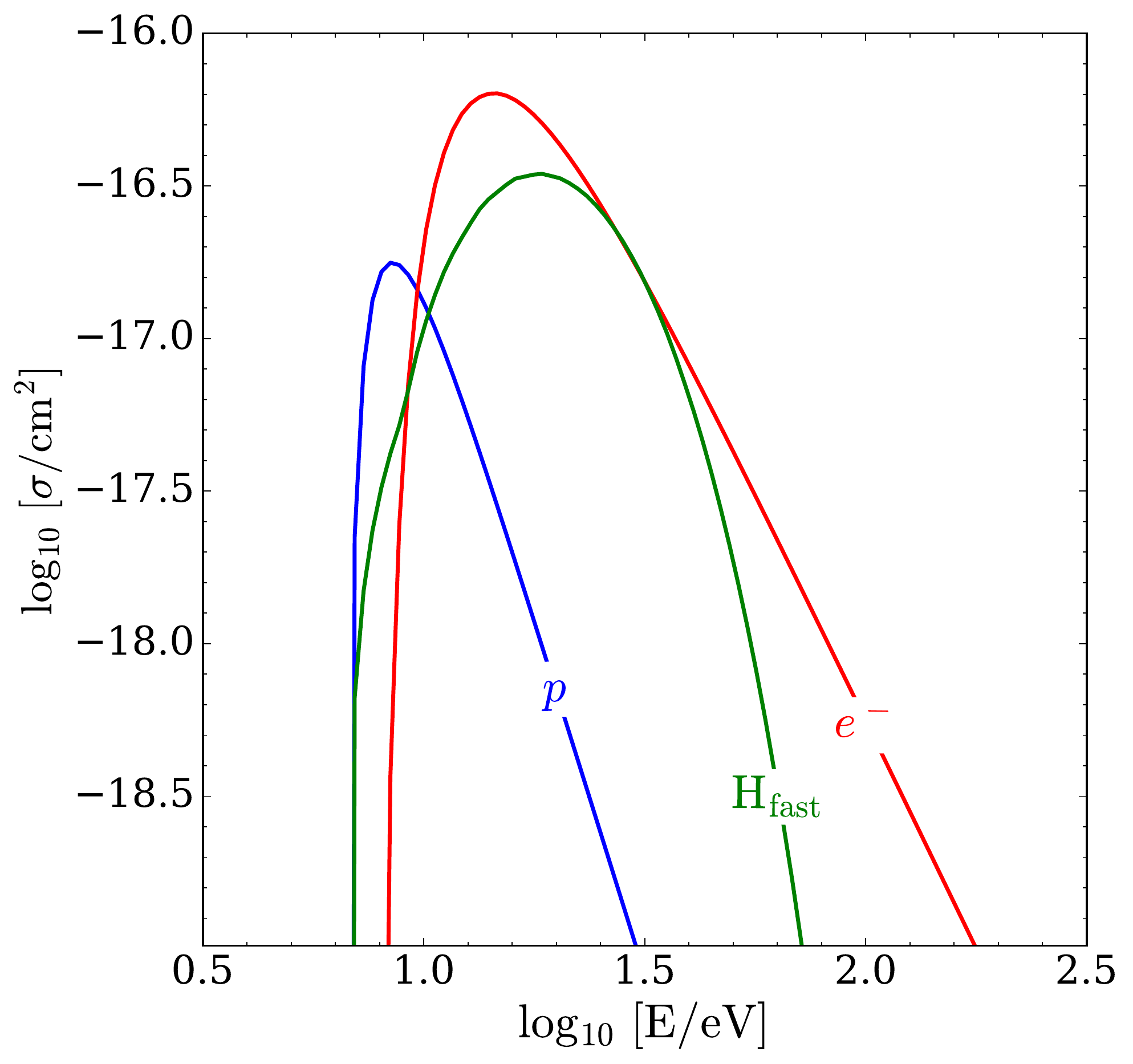}}
\caption{The energy dependence of the dissociation cross sections by protons ({\em blue}), electrons ({\em red}),
and fast hydrogen atoms ({\em green}) colliding with molecular hydrogen.}
\label{fig3}
\end{center}
\end{figure}

Figure~\ref{fig3} shows that dissociation cross sections peak at very low energy, about 8 and 15~eV for protons and
electrons, respectively, so one has to carefully look into the processes that regulate the distributions of different
species in this energy range. In Appendix~\ref{app:equilibrium} we demonstrate that CR protons are efficiently
neutralised at low energies because of electron capture (see also~\citealt{c16}). This generates a flux of fast H
atoms (hereafter H$_{\rm fast}$) that, in turn, creates fast $\h^{+}$ ions (``secondary'' CR protons) through
the reaction (\ref{sfast}). We compute the equilibrium distributions of protons and $\hf$ atoms, finding that below
$\approx10^{4}$~eV less than 10\% of (non-molecular) hydrogen is in the form of $\h^{+}$ (see Fig.~\ref{fq}
in Appendix~\ref{app:equilibrium}), so that the dissociation by H$_{\rm fast}$ (reaction~\ref{sdisshh2}) must be
taken into account. The corresponding cross section, $\sigma_{\rm diss}^{\rm H}$
\citep{dm86,ec09} is also plotted in Fig.~\ref{fig3}.

\subsection{CR dissociation rate}\label{crdissrate}

The rate of dissociation due to primary and secondary CRs and H$_{\rm fast}$ atoms, occurring at the total column
density $N$, is given by
\begin{equation}\label{totaldiss}
\zeta_{\rm diss}^{k}(N)=2\pi\ell\int j_{k}(E,N)\sigma^{k}_{\rm diss}(E)\ud E\,,
\end{equation}
where $j_{k}$ is the differential flux of CR particles $k$, $\sigma_{\rm diss}^{k}$ is the dissociation cross section, and
$k=p,e,{\rm H_{fast}}$. In the semi-infinite slab geometry, the factor $\ell$ is equal to 1 for primary CRs and
H$_{\rm fast}$, and to 2 for secondary electrons (because the latter are produced isotropically). The final
expression for the dissociation rate is obtained by averaging over the pitch-angle distribution of the incident CRs (see
Eq.~45 in~\citealt{pi18}).

\begin{figure}[!h]
\begin{center}
\resizebox{\hsize}{!}{\includegraphics{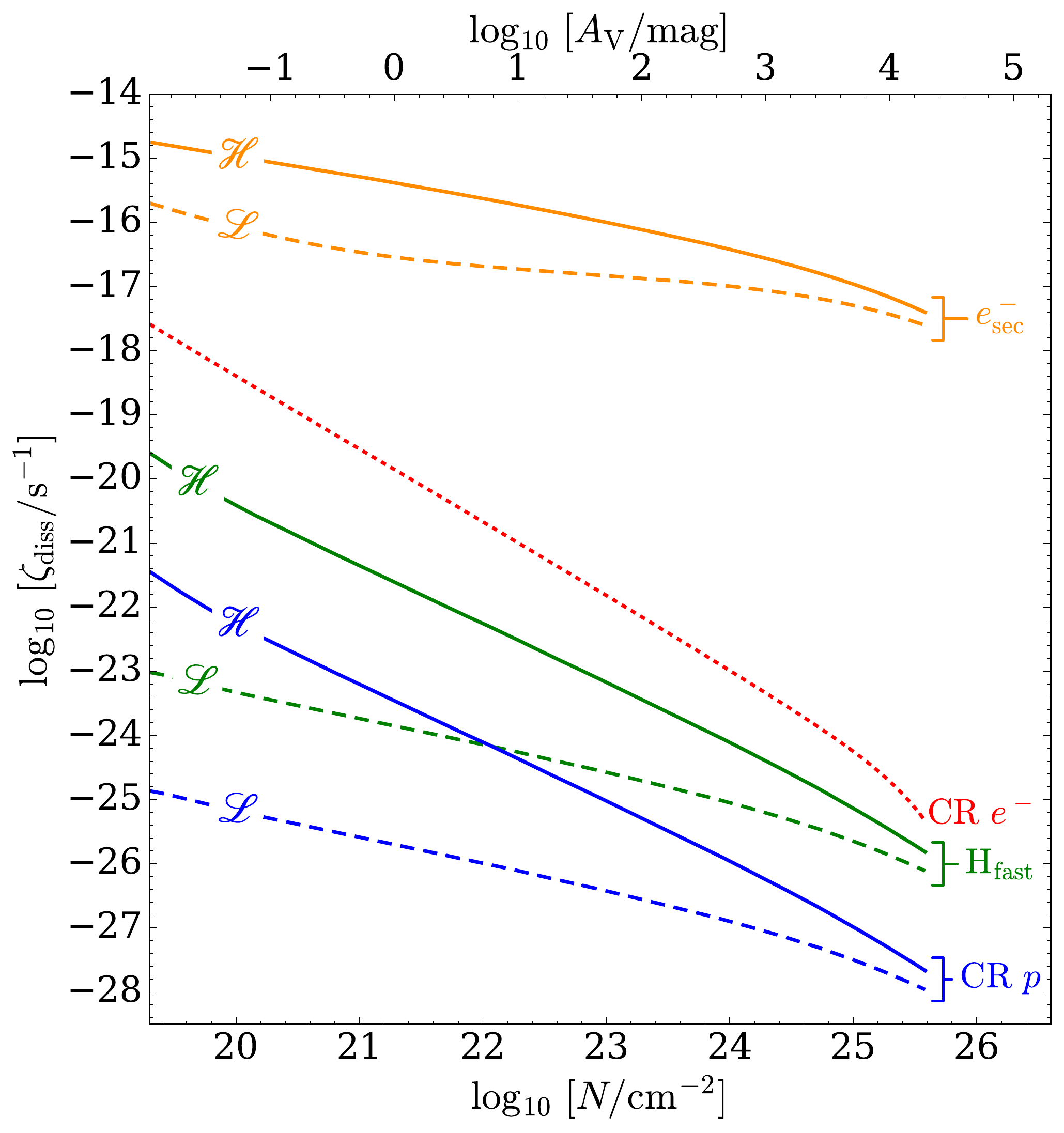}}
\caption{CR dissociation rate for models $\mathscr{L}$ and $\mathscr{H}$ ({\em dashed} and {\em solid lines}, respectively)
as a function of the total column density of hydrogen
(bottom scale)
and visual extinction (top scale). The contributions of primary CR protons
({\em blue}) and electrons ({\em red dotted}), secondary electrons ({\em orange}), and fast H atoms ({\em green})
are shown.}
\label{fig1}
\end{center}
\end{figure}

In the following we assume the same IS CR proton and electron spectra as in~\citet{ip15}
and~\citet{pi18}. For CR protons we adopt two different models: the first one, model $\mathscr{L}$, is an extrapolation of
the Voyager~1 observations to lower energies; the second one, model $\mathscr{H}$, is characterised by an enhanced flux of
low-energy protons with respect to Voyager~1 data. Models $\mathscr{L}$ and $\mathscr{H}$ can be regarded as the lower and
the upper bound, respectively, of the average Galactic CR proton spectrum, since the corresponding CR ionisation rates
encompass the values estimated from observations in diffuse clouds (e.g.~\citealt{in15,nw17}). For CR electrons, we
use a single model based on the latest Voyager results, which show that the electron flux varies at $E\lesssim100$~MeV as
$\propto E^{-1.3}$~\citep{cs16}. Figure~\ref{fig1} shows the partial contributions to the dissociation rate of primary CR
protons and electrons, H$_{\rm fast}$ atoms, and secondary electrons. The latter is computed following Eq.~(16)
in~\citet{ip15}. In Fig.~\ref{fig1} we also show the corresponding visual extinction, $A_{\rm
V}=5.32\times10^{-22}~(N/\mathrm{cm}^{-2})$. One can see that $\zeta_{\rm diss}$ is entirely dominated by low-energy
secondary electrons, produced during the propagation of primary CRs.

\begin{figure}[!h]
\begin{center}
\resizebox{\hsize}{!}{\includegraphics{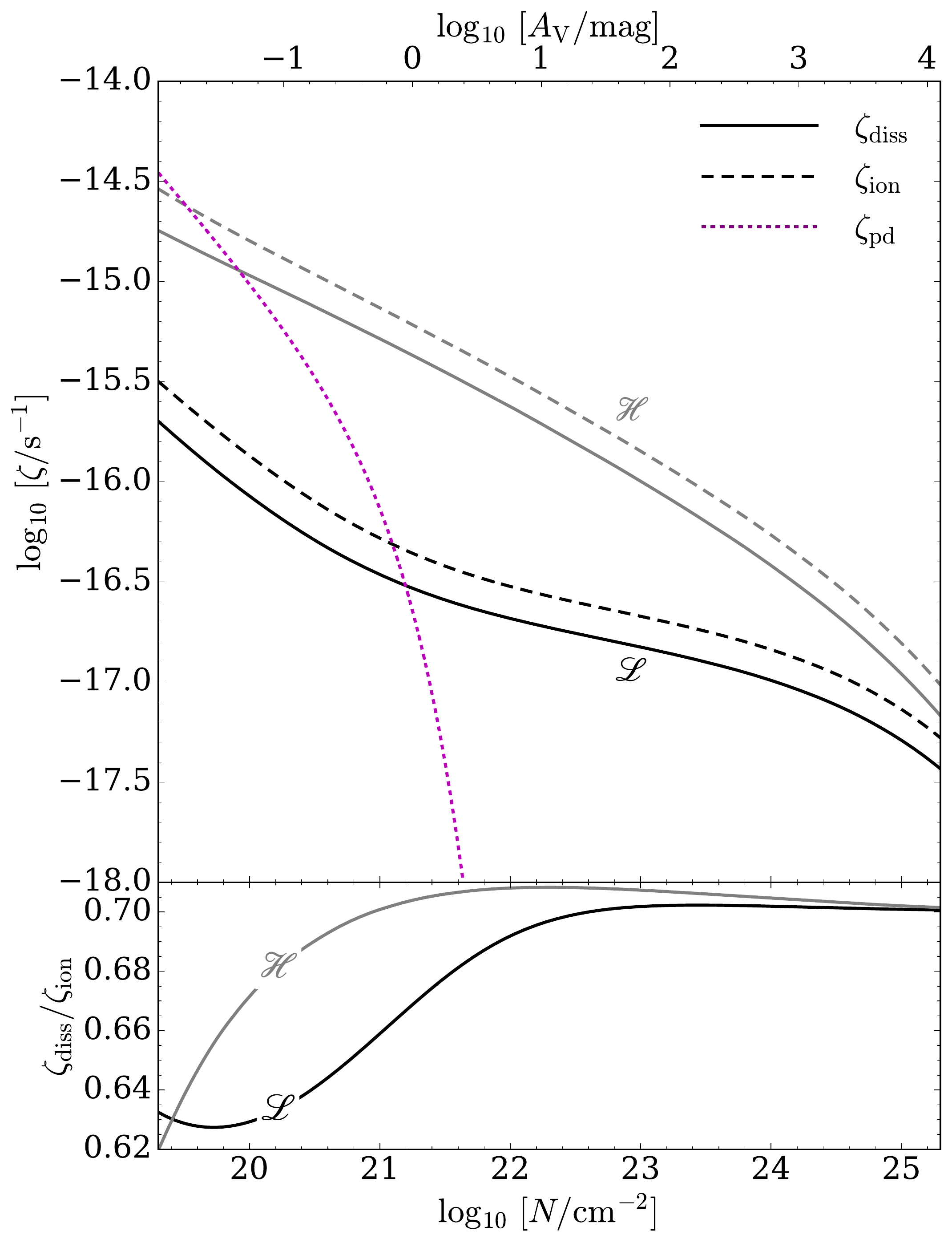}}
\caption{Upper panel: rates of CR dissociation ($\zeta_{\rm diss}$, {\em solid lines}), CR ionisation
($\zeta_{\rm ion}$, {\em dashed lines}), and photodissociation ($\zeta_{\rm pd}$, {\em purple dotted line}) for models
$\mathscr{L}$ ({\em black}) and $\mathscr{H}$ ({\em grey}), plotted versus the total column density of hydrogen
(bottom scale)
and visual extinction (top scale). Lower panel:
ratio $\zeta_{\rm diss}/\zeta_{\rm ion}$ for the two models.}
\label{fig2}
\end{center}
\end{figure}

In previous work \citep[e.g.,][]{lg03,gl05}, $\zeta_{\rm diss}$ has usually been assumed to be equal to the CR ionisation
rate, $\zeta_{\rm ion}$ (which, in turn, did not depend on $N$). In Fig.~\ref{fig2} we show that $\zeta_{\rm
diss}$ and $\zeta_{\rm ion}$ exhibit very similar behaviour, decreasing monotonically with $N$; the ratio $\zeta_{\rm
diss}/\zeta_{\rm ion}$ can be as small as $\approx0.63$ at low column densities ($N\approx10^{19}$~cm$^{-2}$), depending on
the assumed spectrum of IS CR protons. This ratio rapidly approaches the constant value of $\approx0.7$, and at
$N\gtrsim10^{22}$~cm$^{-2}$ becomes independent of the column density and the IS proton spectrum. The values of $\zeta_{\rm
diss}$ and $\zeta_{\rm ion}$ are comparable because secondary electrons provide the major contribution to both
processes. We note that
the ionisation rate has been computed by taking into account the presence of $\hf$ atoms (see Eq.~\ref{crionrate} in
Appendix~\ref{app:ionisation}), contributing to the production of $\hh^{+}$ ions through the reaction~(\ref{sslow}) at
energies below $\approx10^{4}$~eV. However this process is only marginally important for model $\mathscr{H}$ below
$N\approx10^{21}$~cm$^{-2}$, and is always negligible for model $\mathscr{L}$ (see Appendix~\ref{app:ionisation}).
Figure~\ref{fig2} also shows the photodissociation rate, $\zeta_{\rm pd}=D_{0}\chi_{\rm a}$, computed following
\citet{dbook}.

\section{Balance equation}\label{baleq}

\citet{gl05} and \citet{gl07} presented a time-dependent modelling of the H abundance in molecular clouds and
introduced the concept of atomic-to-molecular hydrogen ratio, $n_\h/n_\hh$, as a clock of the cloud's evolutionary stage. In
particular, \citet{gl05} modelled observations of $n_\h/n_\hh$ in five dark clouds, concluding that the
characteristic time required to reach a steady-state $n_\h/n_\hh$ ratio is close to the cloud ages. 
In the following, we consider the steady-state
solution, keeping in mind that time dependence may still affect the interpretation of the observational data (see
Sect.~\ref{uncert}).

In steady-state, the balance between $\hh$ formation and destruction processes gives
\begin{equation}\label{balance}
Rnn_{\h}=n_{\hh}\left(D_{0}\chi_{\rm a}+\zeta_{\rm diss}\right)\,.
\end{equation}
Here, $n=n_{\h}+2n_{\hh}$ is the total volume density of hydrogen, $R$ is the $\hh$ formation rate coefficient,
$D_{0}$ is the unattenuated photodissociation rate, $\chi_{\rm a}$ is the attenuation factor for dust absorption and
$\hh$-self shielding, and $\zeta_{\rm diss}$ is the CR dissociation rate. In the following we assume
$R=3\times10^{-17}$~cm$^{3}$~s$^{-1}$ \citep{j75} and $D_{0}=2\times10^{-11}G_{0}$~s$^{-1}$ \citep[][taking into
account a semi-infinite slab geometry]{dbook}, where $G_{0}$ is the FUV radiation field in Habing units \citep{h68}. Unless
specified otherwise, we adopt $G_{0}=1$. The attenuation factor is usually written in the form
\begin{equation}
\chi_{\rm a}(N,N_{\hh})=\chi_{\rm sh}(N_{\hh})e^{-\tau(N)}\,,
\end{equation}
where $\chi_{\rm sh}(N_{\hh})=(10^{14}~\mathrm{cm^{-2}}/N_{\hh})^{0.75}$ is the $\hh$ self-shielding factor \citep[][valid
for $10^{14}~{\rm cm}^{-2}\lesssim N_{\hh}\lesssim 10^{21}~{\rm cm}^{-2}$]{tielensbook} and $\tau(N)=\sigma_{\rm g}N$ is the
dust attenuation. Here, $N=N_{\h}+2N_{\hh}$ is the total column density of hydrogen and $\sigma_{\rm
g}=1.9\times10^{-21}$~cm$^{2}$ is the average value of the FUV dust grain absorption cross section for solar
metallicity \citep{dbook}.

Assuming $n_{\h}/n=\ud N_{\h}/\ud N$ and $n_{\hh}/n=\ud N_{\hh}/\ud N$, Eq.~(\ref{balance}) becomes
\begin{equation}\label{diffeq}
\frac{\ud N_{\hh}}{\ud N}=\left(2+\frac{D_{0}\chi_{\rm a}+\zeta_{\rm diss}}{Rn}\right)^{-1}\,.
\end{equation}
The fractions of atomic and molecular hydrogen can be expressed as
\begin{equation}\label{fh}
f_{\h}=\frac{n_{\h}}{n_{\h}+n_{\hh}}=\frac{1-2\ud N_{\hh}/\ud N}{1-\ud N_{\hh}/\ud N}
\end{equation}
and
\begin{equation}\label{fhh}
f_{\hh}=1-f_{\h}=\frac{\ud N_{\hh}/\ud N}{1-\ud N_{\hh}/\ud N}\,,
\end{equation}
respectively. In the next section we describe in detail all the processes that contribute to the dissociation of molecular
hydrogen.

\section{Comparison with observations}\label{comparison}

\citet{lg03} performed a survey of dark clouds in the Taurus-Perseus region, and reported the detection of \HI\ narrow
self-absorption features. This allowed them to compute the atomic and molecular hydrogen fraction (Eqs.~\ref{fh} and
\ref{fhh}). They concluded that a relevant fraction of atomic hydrogen is mixed with $\hh$ in the densest part of a cloud,
shielded from the IS UV flux. 

At high column densities typical of dark clouds, the attenuation factor $\chi_{\rm a}$ in Eq.~(\ref{balance}) is so large
that the UV photodissociation is inefficient, and the observed $n_\h/n_\hh$ ratios can only be explained by CR dissociation.
In Sect.~\ref{crdissrate}, we showed that $\zeta_{\rm diss}\approx0.7\zeta_{\rm ion}$ at typical column densities of dark
clouds ($\approx10^{22}$~cm$^{-2}$); more importantly, $\zeta_{\rm diss}$ is not constant, but decreases with $N$
(e.g., \citealt{pgg09,pi18}).

We compute the fraction of atomic and molecular hydrogen expected at different column densities (Eqs.~\ref{fh} and
\ref{fhh}), to evaluate the effect of CR dissociation on the abundance of atomic hydrogen in dark clouds. For
the total volume density $n$ in Eq.~(\ref{balance}), we use the average value of $5\times10^{3}$~cm$^{-3}$
computed by \citet{lg03}, to which we add an error of $2.6\times10^{3}$~cm$^{-3}$ (the standard deviation for the
observed values).

\begin{figure}[!h]
\begin{center}
\resizebox{\hsize}{!}{\includegraphics{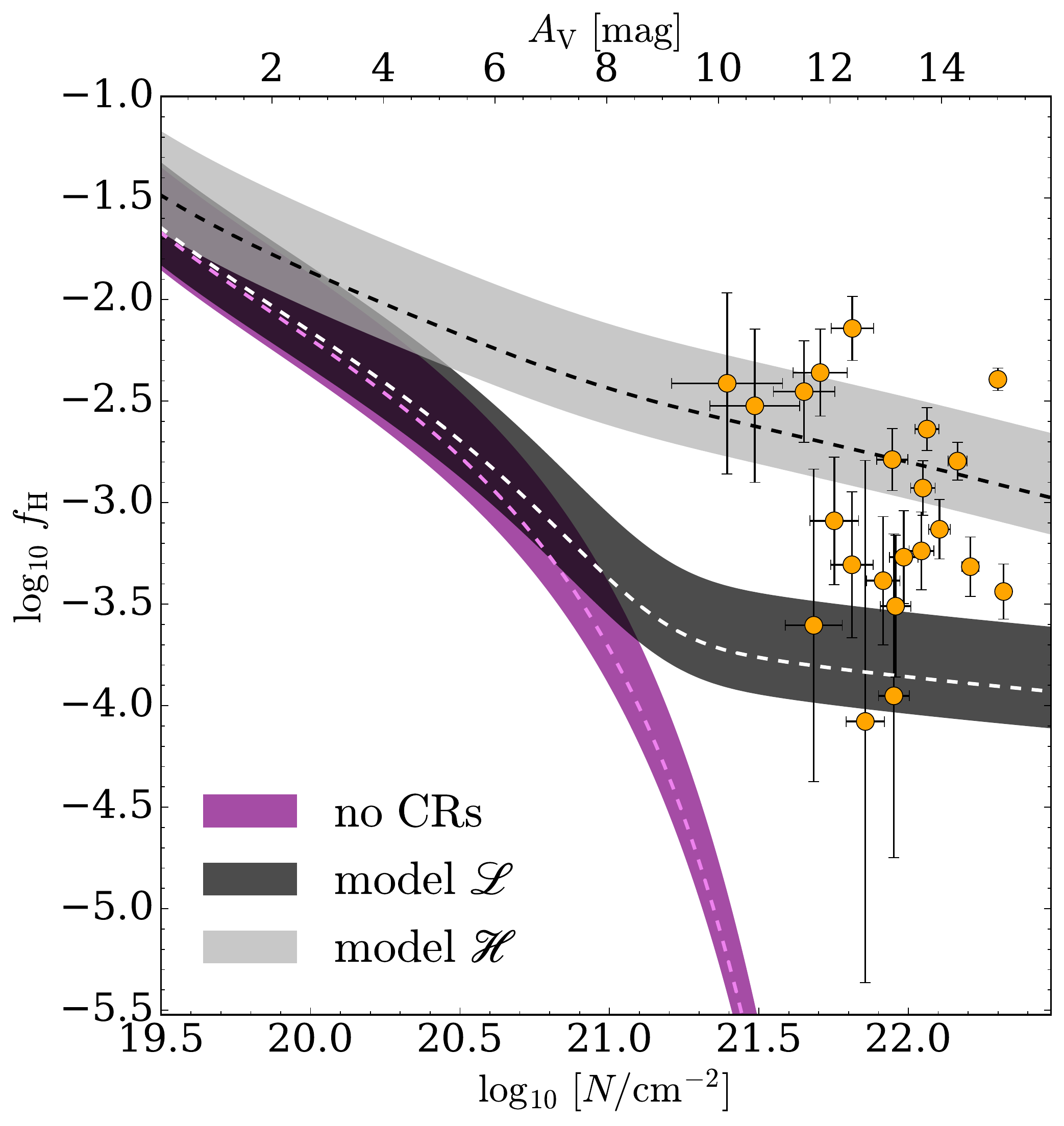}}
\caption{Atomic hydrogen fraction versus the total column density of hydrogen (bottom scale)
and visual extinction (top scale).
Observations from \citet{lg03} are shown as {\em solid orange circles}. {\em Coloured stripes}
represent our results for the case of photodissociation only
({\em purple}), model $\mathscr{L}$ ({\em black}), and $\mathscr{H}$ ({\em grey}).
{\em Dashed lines} refer to the average value of the total volume density of hydrogen \cite[suggested by][]{lg03}.}
\label{fig4}
\end{center}
\end{figure}

Figure~\ref{fig4} shows the comparison between our models and the observations by \citet{lg03}.
As expected, UV photodissociation alone cannot explain the observed $n_\h/n_\hh$ ratios because of the attenuation
at large column densities. More notably, a CR spectrum based on the extrapolation of the Voyager data (model
$\mathscr{L}$) fails to reproduce the majority of the observations, and only a spectrum enhanced at low energies
(such as model $\mathscr{H}$) can explain this. The latter fact corroborates the need of a low-energy tail in
the IS CR flux of protons, also required to explain the high CR ionisation rates in diffuse clouds
(e.g.~\citealt{pgg09};~\citealt{in15}).

\subsection{Uncertainties of the $\hh$ formation rate}\label{uncert}

The large spread in the observed values of $f_{\h}$ probably reflects a broad variety of environments in dark clouds,
including, e.g., variations in the density and IS UV radiation field (see e.g.~\citealt{bs16}). In this work we assume a
$\hh$ formation rate of $R=3\times10^{-17}$~cm$^{3}$~s$^{-1}$ (see Sect.~\ref{baleq})%
\footnote{We note that \citet{lg03} used $R=6.5\times10^{-18}$~cm$^{3}$~s$^{-1}$, which is a factor of $\approx$5 smaller
than our value. A lower $R$ implies a lower ionisation rate needed to reproduce the observations. This explains why, using a
constant dissociation rate equal to the ionisation rate of $3\times10^{-17}$~s$^{-1}$, they found $f_{\rm
H}\approx1.5\times10^{-3}$.}. However, $R$ is strongly dependent on the condition of each cloud; for example, in
photodissociation regions, where the large abundance of polycyclic aromatic hydrocarbons favours the formation of $\hh$, $R$
can increase by one order of magnitude \citep{hb04}. Variations in the grain size distribution may also change the value of
$R$ by a factor of $\sim3$ \citep{gl05}. \citet{dbook} suggests $R\approx3\times10^{-17}\sqrt{T/70~{\rm
K}}$~cm$^{3}$~s$^{-1}$, but even assuming $T$ as low as 10~K, we find $f_{\rm H}\approx10^{-3}$ at
$N\approx10^{22}$~cm$^{-2}$ for a Voyager-like spectrum (model $\mathscr{L}$). As a consequence, a larger flux of
low-energy CR protons (model $\mathscr{H}$) is still needed to explain the higher $n_\h/n_\hh$ ratios. This conclusion
remains unchanged even if $G_{0}$ is increased by up to two orders of magnitude, since the UV field is exponentially
attenuated in the range of column densities of the observed dark clouds ($2\times10^{21}~{\rm cm^{-2}}\lesssim
N\lesssim2\times10^{22}~{\rm cm^{-2}}$).

One should also keep in mind that, as mentioned in Sect.~\ref{baleq}, some of the observed clouds may have not
necessarily reached a steady-state $n_{\h}/n_{\hh}$ ratio~\citep{gl05,gl07}. In this case, model $\mathscr{L}$ could not be
completely ruled out as, if these clouds are younger, they will have higher $n_{\h}/n_{\hh}$ ratio than predicted by our
steady-state assumption. 

\section{Discussion and conclusions}\label{discconc}

Dissociation of H$_2$ into atomic hydrogen by CRs in dark clouds can have important consequences for the
chemical evolution of dense regions in the clouds. Atomic hydrogen is the most mobile reactive species
on the surface of bare dust grains and icy mantles, and therefore it is crucial to accurately determine its abundance.
The larger $f_{\h}$ values predicted in this work imply a more efficient hydrogenation of molecular species
on grain surfaces.
In particular, hydrogenation of CO, which freezes out onto grains at densities above a few
$10^{4}$~cm$^{-3}$ (e.g., \citealt{cw99}), follows the sequence
(e.g. \citealt{th82})
\begin{equation}\label{cohyd}
{\rm CO}\xrightarrow{\rm H}{\rm HCO}\xrightarrow{\rm H}{\rm H_{2}CO}\xrightarrow{\rm H}{\rm H_{3}CO}\xrightarrow{\rm
H}{\rm CH_{3}OH}\,.
\end{equation}
This leads to efficient production of formaldehyde (H$_{2}$CO) and methanol (CH$_{3}$OH; \citealt{vc17})
on very short time scales.
Hence, even if dissociation by energetic particles takes place, CO cannot be returned to the gas phase, because
it is rapidly converted into methanol.
If the products of dissociation do not move very far from their formation site
\citep{st18}, methanol is ejected from the surface.
This is because the exothermicity of chemical reactions (\ref{cohyd}) is partially channeled into kinetic energy through a process known as reactive desorption \citep{gw07}.
On the other hand, ammonia (NH$_3$), which is synthesised onto grains through the hydrogenation
sequence (e.g. \citealt{hy95,fi15})
\be
{\rm N}\xrightarrow{\rm H}{\rm NH}\xrightarrow{\rm H}{\rm NH_2}\xrightarrow{\rm H}{\rm NH_3}\,,
\ee
can, in principle, go back to the gas phase upon surface dissociation followed by reactive desorption.
These considerations could
help in explaining the observational evidence that NH$_3$ (unlike CO) does not appear to deplete towards the central
regions of dense cores, despite its large binding energy.
To verify this hypothesis, one should carefully evaluate the consequences of an enhanced abundance of
atomic hydrogen
in chemical models.

We point out that CR dissociation is not only limited to $\hh$, but could occur for other molecular species as
well, both in the gas phase and on/in ices mantles, with potentially significant consequences
in the chemical composition of dense cloud cores and
dark clouds. 



To summarise, in this paper
we studied the role of CRs in determining the fractional abundance of atomic hydrogen in dark
clouds. The main results are:
\begin{itemize}
\item[($i$)] The CR dissociation rate, $\zeta_{\rm diss}$, is primarily
	determined by secondary electrons produced during
    the primary CR ionisation process. These secondary electrons can efficiently dissociate H$_2$ and represent
    the only source of atomic hydrogen at column densities larger than $\approx10^{21}$~cm$^{-2}$,
    regulating the $n_\h/n_\hh$ ratio in dark clouds;
\item[($ii$)] $\zeta_{\rm diss}$ entering the balance equation (\ref{balance}) is not equal to the
    ionisation rate $\zeta_{\rm ion}$, as assumed in some previous work. We find that the ratio $\zeta_{\rm
    diss}/\zeta_{\rm ion}$ varies between $\approx0.63$ and $\approx0.7$, depending on the column density range,
    while $\zeta_{\rm ion}$ is a decreasing function of the column density;
\item[($iii$)] Even given
	the uncertainties in the values of $\hh$ formation rate, temperature, total hydrogen volume
    density, and IS UV radiation field for each cloud, only a CR proton spectrum enhanced at low energies (such as our
    model $\mathscr{H}$) is capable to reproduce the upper values of measured $f_{\rm H}$,
    under the assumption of steady state. We note that
    neither model $\mathscr{L}$ nor $\mathscr{H}$ is able to reproduce the entire set of observational data:
    the spread in
    the values of $f_{\rm H}$ at any given column density must be attributed to
    time dependence or to individual characteristics of each cloud.
    For example, tangled magnetic field lines and/or higher volume densities would result in a stronger CR attenuation and
    therefore in a lower $f_{\rm H}$;
\item[($iv$)] An accurate description of H$_2$ dissociation in dense environments is essential, because many
    chemical processes (such as CO hydrogenation and its depletion degree onto dust grains,
    or formation of complex organic molecules) critically depend on the abundance of atomic hydrogen.
\end{itemize}

\begin{acknowledgements}
The authors wish to thank the referee, Paul Goldsmith,
for his careful reading of the manuscript and insightful comments that considerably helped to improve the paper.
MP acknowledges funding from the European Union's Horizon 2020 research and innovation programme under the Marie
Sk\l{}odowska-Curie grant agreement No 664931.
AF acknowledges support from the ERC Advanced Grant INTERSTELLAR H2020/740120.
The authors thank Fabrizio Esposito for sending us his results for the $\hh$ dissociation cross section
by atomic hydrogen impact.
\end{acknowledgements}

\appendix
\section{Equilibrium distribution of protons and $\hf$ atoms at low energies}\label{app:equilibrium}

Because of the process of electron capture at low energies, CR protons interacting with $\hh$ are efficiently
neutralised,
\begin{equation}\label{sec}
p+{\rm H}_{2}\rightarrow {\rm H}_{2}^{+}+{\rm H}_{\rm fast}\quad(\sigma^{p}_{\rm e.c.})\,,
\end{equation}
creating fast H atoms\footnote{In parentheses, we put the cross section of the respective process.}. At the
same time, H$_{\rm fast}$ atoms reacting with H$_{2}$ yield
\begin{eqnarray}
{\rm H}_{\rm fast}+{\rm H}_{2} &\rightarrow& \mathrm{H_{\rm fast}+H_{2}^{+}}+e\quad(\sigma^{\rm H}_{\rm ion})\,;\label{sslow}\\
&& p+{\rm H_{2}}+e\;\quad\quad(\sigma^{\rm H}_{\rm self-ion})\label{sfast}\,;\\
&& {\rm H}_{\rm fast}+\mathrm{H+H}\,\quad(\sigma^{\h}_{\rm diss})\,.\label{sdisshh2}
\end{eqnarray}
The electron capture by protons, Eq.~(\ref{sec}), and the reaction of $\hf$ self-ionisation, Eq.~(\ref{sfast}), are
catastrophic processes, since the respective projectile particles disappear after such collisions. The reactions of $\hh$
ionisation and dissociation by $\hf$ atoms, Eqs.~(\ref{sslow}) and (\ref{sdisshh2}), respectively, are continuous loss
processes, where the projectile kinetic energy decreases only slightly
after each collision. The efficiency of continuous
energy losses is generally characterised by the projectile's stopping range (see, e.g.~\citealt{pgg09}).

\begin{figure}[]
\begin{center}
\resizebox{\hsize}{!}{\includegraphics{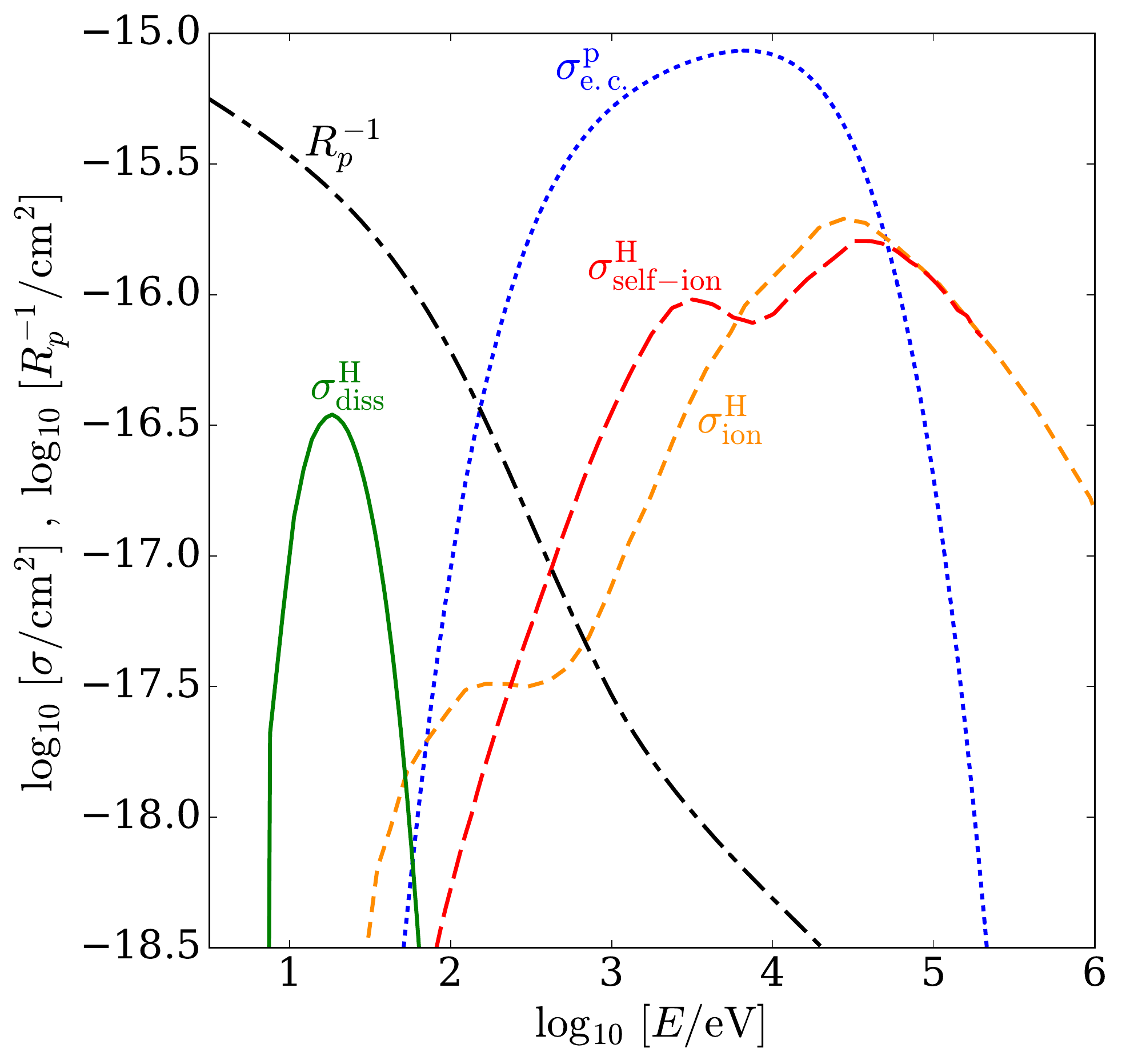}}
\caption{The cross sections of processes governing equilibrium distributions of protons and fast hydrogen atoms at
low energies: electron capture by $p$ (\ref{sec}, {\em dotted blue line}), ionisation of $\hh$ by $\hf$ (\ref{sslow}, {\em
short-dashed orange line}), self-ionisation of $\hf$ (\ref{sfast}, {\em long-dashed red line}), and $\hh$ dissociation by
$\hf$ (\ref{sdisshh2}, {\em solid green line}). The inverse of the proton stopping range, $R_{p}^{-1}$, is also plotted
{(\em black dot-dashed line)}.} \label{allsigma}
\end{center}
\end{figure}

\begin{figure}[]
\begin{center}
\resizebox{\hsize}{!}{\includegraphics{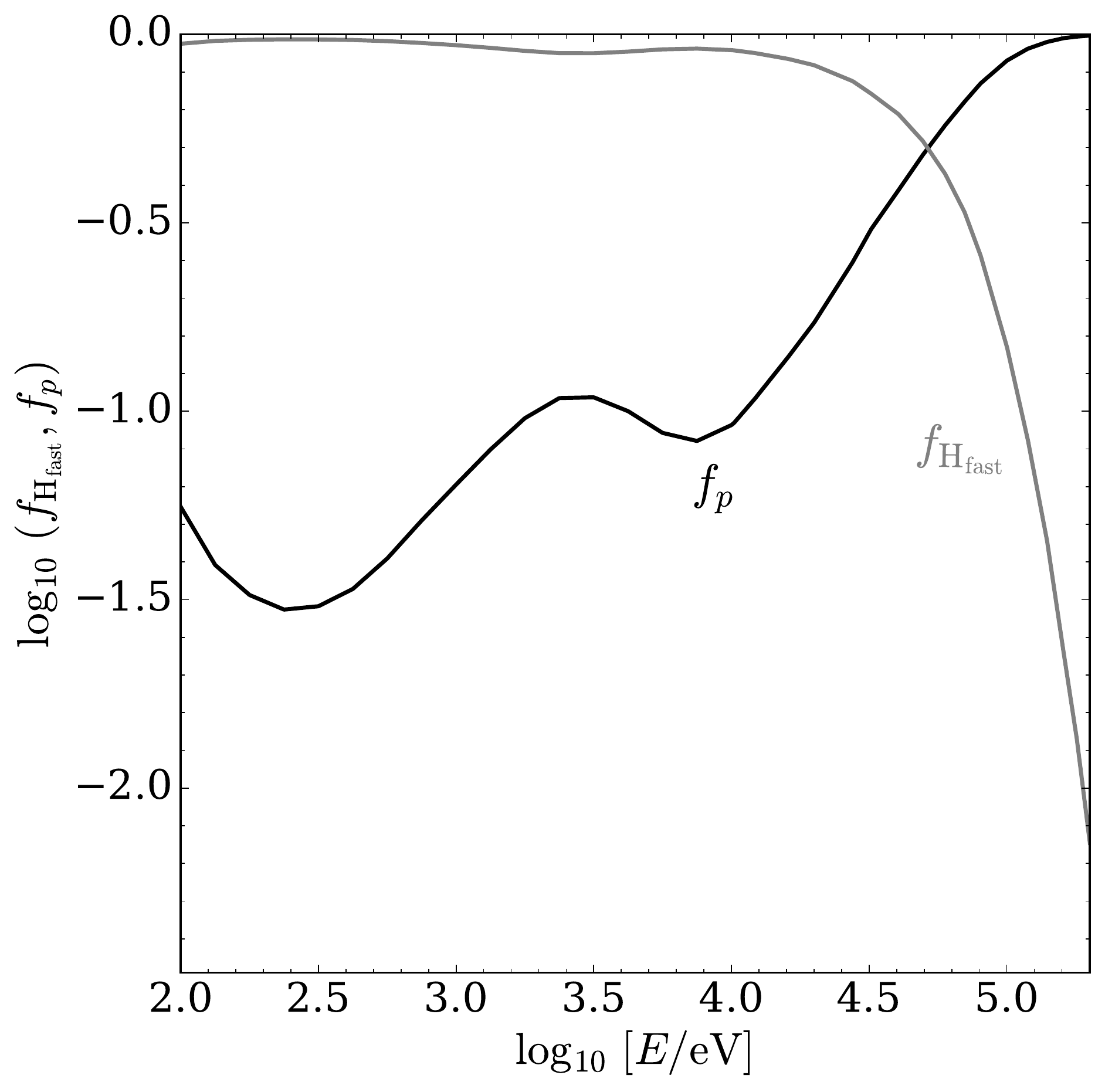}}
\caption{Fraction of non-molecular hydrogen in neutral ($f_{\hf}$) and ionised ($f_{p}$) form as a function of the
energy.}
\label{fq}
\end{center}
\end{figure}

For our calculations, $\sigma^{p}_{\rm e.c.}$ is taken from \citet{rg83}, $\sigma^{\h}_{\rm ion}$ is from \citet{p90}
and \citet{ks91}, $\sigma^{\h}_{\rm self-ion}$ is computed by \citet{s56}, \citet{vla81}, and \citet{p90}, and
$\sigma^{\h}_{\rm diss}$ is from \citet{dm86} and \citet{ec09}. In Fig.~\ref{allsigma} we plot the cross sections and the inverse of the
proton stopping range, $R_p^{-1}$, versus the respective projectile's energy. We see that $\sigma^{p}_{\rm e.c.}$ is much
larger than $R_p^{-1}$ for $10^{2}~{\rm eV}\lesssim E\lesssim 10^{5}~{\rm eV}$, which implies that continuous loss processes
cannot significantly affect the balance between protons and $\hf$ atoms at these energies. The equilibrium ratio of the
$\hf$ and proton fluxes is then given by
\begin{equation}\label{jhpjh}
\frac{j_{\hf}}{j_p}\approx\frac{\sigma^{p}_{\rm e.c.}}{\sigma^{\h}_{\rm self-ion}}\,.
\end{equation}
This allows us to calculate the fractions of H$_{\rm fast}$ atoms,
\begin{equation}
f_{\hf}=\frac{j_{\hf}}{j_{\hf}+j_p}=\frac{\sigma^{p}_{\rm e.c.}}{\sigma^{p}_{\rm e.c.}+\sigma^{\h}_{\rm self-ion}}\,,
\end{equation}
and protons, $f_p=1-f_{\hf}$. Figure~\ref{fq} shows that for energies below $\approx10^{4}$~eV, only less than 10\% of
non-molecular hydrogen is in the form of protons. This means that $\hh$ ionisation at these energies is dominated by H$_{\rm
fast}$ atoms, via reaction~(\ref{sslow}).

\section{Ionisation by $\hf$ atoms}\label{app:ionisation}

As shown in Appendix~\ref{app:equilibrium}, the ionisation at energies below $\approx10^{4}$~eV is mostly driven by
$\hf$ atoms. To take this effect into account, we use the following expression for the $\hh$ ionisation rate by CR protons:
\begin{eqnarray}\label{crionrate}
\zeta_{\rm ion}^{p}(N)&=&2\pi\int \bigg\{j_p(E,N)\left[\sigma^{p}_{\rm ion}(E)+\sigma^{p}_{\rm e.c.}(E)\right]+\\\nonumber
&&j_{\hf}(E,N)\sigma^{\h}_{\rm ion}(E)\bigg\}\ud E\,,
\end{eqnarray}
where $\sigma^{p}_{\rm ion}$ is the $\hh$ ionisation cross section by proton impact (\citealt{rk85}). It turns out,
however, that the difference between the ionisation rates computed from Eq.~(\ref{crionrate}) taking into account
Eq.~(\ref{jhpjh}) and assuming $j_{\hf}=0$ is very small: for $N\approx10^{19}$~cm$^{-2}$, the difference is $\approx5\%$
and $\approx40\%$ for models $\mathscr{L}$ and $\mathscr{H}$, respectively, at higher column densities it rapidly decreases
and becomes negligible for both models above $\approx10^{21}$~cm$^{-2}$. This result justifies the assumption
$j_{\hf}=0$ made previously for calculating the ionisation.

\bibliographystyle{aa} 
\bibliography{mybibliography} 

\begin{thebibliography}{41}
\expandafter\ifx\csname natexlab\endcsname\relax\def\natexlab#1{#1}\fi

\bibitem[{{Bialy} \& {Sternberg}(2016)}]{bs16}
{Bialy}, S. \& {Sternberg}, A. 2016, \apj, 822, 83

\bibitem[{{Caselli} {et~al.}(1999){Caselli}, {Walmsley}, {Tafalla}, {Dore}, \&
  {Myers}}]{cw99}
{Caselli}, P., {Walmsley}, C.~M., {Tafalla}, M., {Dore}, L., \& {Myers}, P.~C.
  1999, \apjl, 523, L165

\bibitem[{{Chabot}(2016)}]{c16}
{Chabot}, M. 2016, \aap, 585, A15

\bibitem[{{Cummings} {et~al.}(2016){Cummings}, {Stone}, {Heikkila}, {Lal},
  {Webber}, {J{\'o}hannesson}, {Moskalenko}, {Orlando}, \& {Porter}}]{cs16}
{Cummings}, A.~C., {Stone}, E.~C., {Heikkila}, B.~C., {et~al.} 2016, \apj, 831,
  18

\bibitem[{{Dove} \& {Mandy}(1986)}]{dm86}
{Dove}, J.~E. \& {Mandy}, M.~E. 1986, \apjl, 311, L93

\bibitem[{{Draine}(2011)}]{dbook}
{Draine}, B.~T. 2011, {Physics of the Interstellar and Intergalactic Medium}

\bibitem[{{Esposito} \& {Capitelli}(2009)}]{ec09}
{Esposito}, F. \& {Capitelli}, M. 2009, Journal of Physical Chemistry A, 113,
  15307

\bibitem[{{Fedoseev} {et~al.}(2015){Fedoseev}, {Ioppolo}, {Zhao}, {Lamberts},
  \& {Linnartz}}]{fi15}
{Fedoseev}, G., {Ioppolo}, S., {Zhao}, D., {Lamberts}, T., \& {Linnartz}, H.
  2015, \mnras, 446, 439

\bibitem[{{Garrod} {et~al.}(2007){Garrod}, {Wakelam}, \& {Herbst}}]{gw07}
{Garrod}, R.~T., {Wakelam}, V., \& {Herbst}, E. 2007, \aap, 467, 1103

\bibitem[{{Goldsmith} \& {Li}(2005)}]{gl05}
{Goldsmith}, P.~F. \& {Li}, D. 2005, \apj, 622, 938

\bibitem[{{Goldsmith} {et~al.}(2007){Goldsmith}, {Li}, \& {Kr{\v c}o}}]{gl07}
{Goldsmith}, P.~F., {Li}, D., \& {Kr{\v c}o}, M. 2007, \apj, 654, 273

\bibitem[{{Habart} {et~al.}(2004){Habart}, {Boulanger}, {Verstraete},
  {Walmsley}, \& {Pineau des For{\^e}ts}}]{hb04}
{Habart}, E., {Boulanger}, F., {Verstraete}, L., {Walmsley}, C.~M., \& {Pineau
  des For{\^e}ts}, G. 2004, \aap, 414, 531

\bibitem[{{Habing}(1968)}]{h68}
{Habing}, H.~J. 1968, \bain, 20, 120

\bibitem[{{Hiraoka} {et~al.}(1995){Hiraoka}, {Yamashita}, {Yachi}, {Aruga},
  {Sato}, \& {Muto}}]{hy95}
{Hiraoka}, K., {Yamashita}, A., {Yachi}, Y., {et~al.} 1995, \apj, 443, 363

\bibitem[{{Hollenbach} {et~al.}(1971){Hollenbach}, {Werner}, \&
  {Salpeter}}]{hw71}
{Hollenbach}, D.~J., {Werner}, M.~W., \& {Salpeter}, E.~E. 1971, \apj, 163, 165

\bibitem[{{Houde} {et~al.}(2009){Houde}, {Vaillancourt}, {Hildebrand},
  {Chitsazzadeh}, \& {Kirby}}]{hh09}
{Houde}, M., {Vaillancourt}, J.~E., {Hildebrand}, R.~H., {Chitsazzadeh}, S., \&
  {Kirby}, L. 2009, \apj, 706, 1504

\bibitem[{{Indriolo} {et~al.}(2015){Indriolo}, {Neufeld}, {Gerin}, {Schilke},
  {Benz}, {Winkel}, {Menten}, {Chambers}, {Black}, {Bruderer}, {Falgarone},
  {Godard}, {Goicoechea}, {Gupta}, {Lis}, {Ossenkopf}, {Persson},
  {Sonnentrucker}, {van der Tak}, {van Dishoeck}, {Wolfire}, \&
  {Wyrowski}}]{in15}
{Indriolo}, N., {Neufeld}, D.~A., {Gerin}, M., {et~al.} 2015, \apj, 800, 40

\bibitem[{{Ivlev} {et~al.}(2015){Ivlev}, {Padovani}, {Galli}, \&
  {Caselli}}]{ip15}
{Ivlev}, A.~V., {Padovani}, M., {Galli}, D., \& {Caselli}, P. 2015, \apj, 812,
  135

\bibitem[{{Janev} {et~al.}(2003){Janev}, {Reiter}, \& {Samm}}]{janevbook}
{Janev}, R.~K., {Reiter}, D., \& {Samm}, U. 2003, {Collision processes in
  low-temperature hydrogen plasmas}, 188

\bibitem[{{Jura}(1975)}]{j75}
{Jura}, M. 1975, \apj, 197, 575

\bibitem[{{Kunc} \& {Soon}(1991)}]{ks91}
{Kunc}, J.~A. \& {Soon}, W.~H. 1991, \jcp, 95, 5738

\bibitem[{{Li} \& {Goldsmith}(2003)}]{lg03}
{Li}, D. \& {Goldsmith}, P.~F. 2003, \apj, 585, 823

\bibitem[{{McCutcheon} {et~al.}(1978){McCutcheon}, {Shuter}, \& {Booth}}]{ms78}
{McCutcheon}, W.~H., {Shuter}, W.~L.~H., \& {Booth}, R.~S. 1978, \mnras, 185,
  755

\bibitem[{{Montgomery} {et~al.}(1995){Montgomery}, {Bates}, \& {Davies}}]{mb95}
{Montgomery}, A.~S., {Bates}, B., \& {Davies}, R.~D. 1995, \mnras, 273, 449

\bibitem[{{Neufeld} \& {Wolfire}(2017)}]{nw17}
{Neufeld}, D.~A. \& {Wolfire}, M.~G. 2017, \apj, 845, 163

\bibitem[{{Padovani} \& {Galli}(2011)}]{pg11}
{Padovani}, M. \& {Galli}, D. 2011, \aap, 530, A109

\bibitem[{{Padovani} \& {Galli}(2013)}]{pg13}
{Padovani}, M. \& {Galli}, D. 2013, in Astrophysics and Space Science
  Proceedings, Vol.~34, Cosmic Rays in Star-Forming Environments, ed. D.~F.
  {Torres} \& O.~{Reimer}, 61

\bibitem[{{Padovani} {et~al.}(2009){Padovani}, {Galli}, \& {Glassgold}}]{pgg09}
{Padovani}, M., {Galli}, D., \& {Glassgold}, A.~E. 2009, \aap, 501, 619

\bibitem[{{Padovani} {et~al.}(2013){Padovani}, {Hennebelle}, \& {Galli}}]{ph13}
{Padovani}, M., {Hennebelle}, P., \& {Galli}, D. 2013, \aap, 560, A114

\bibitem[{{Padovani} {et~al.}(2018){Padovani}, {Ivlev}, {Galli}, \&
  {Caselli}}]{pi18}
{Padovani}, M., {Ivlev}, A.~V., {Galli}, D., \& {Caselli}, P. 2018, ArXiv
  e-prints [\eprint[arXiv]{1803.09348}]

\bibitem[{{Phan} {et~al.}(2018){Phan}, {Morlino}, \& {Gabici}}]{phan18}
{Phan}, V.~H.~M., {Morlino}, G., \& {Gabici}, S. 2018, ArXiv e-prints
  [\eprint[arXiv]{1804.10106}]

\bibitem[{{Phelps}(1990)}]{p90}
{Phelps}, A.~V. 1990, Journal of Physical and Chemical Reference Data, 19, 653

\bibitem[{{Rudd} {et~al.}(1983){Rudd}, {Goffe}, {Dubois}, {Toburen}, \&
  {Ratcliffe}}]{rg83}
{Rudd}, M.~E., {Goffe}, T.~V., {Dubois}, R.~D., {Toburen}, L.~H., \&
  {Ratcliffe}, C.~A. 1983, \pra, 28, 3244

\bibitem[{{Rudd} {et~al.}(1985){Rudd}, {Kim}, {Madison}, \& {Gallagher}}]{rk85}
{Rudd}, M.~E., {Kim}, Y.-K., {Madison}, D.~H., \& {Gallagher}, J.~W. 1985,
  Reviews of Modern Physics, 57, 965

\bibitem[{{Shingledecker} {et~al.}(2018){Shingledecker}, {Tennis}, {Le Gal}, \&
  {Herbst}}]{st18}
{Shingledecker}, C.~N., {Tennis}, J., {Le Gal}, R., \& {Herbst}, E. 2018, \apj,
  861, 20

\bibitem[{{Stier} \& {Barnett}(1956)}]{s56}
{Stier}, P.~M. \& {Barnett}, C.~F. 1956, Physical Review, 103, 896

\bibitem[{{Tielens}(2010)}]{tielensbook}
{Tielens}, A.~G.~G.~M. 2010, {The Physics and Chemistry of the Interstellar
  Medium}

\bibitem[{{Tielens} \& {Hagen}(1982)}]{th82}
{Tielens}, A.~G.~G.~M. \& {Hagen}, W. 1982, \aap, 114, 245

\bibitem[{{van der Werf} {et~al.}(1988){van der Werf}, {Goss}, \& {Vanden
  Bout}}]{wg88}
{van der Werf}, P.~P., {Goss}, W.~M., \& {Vanden Bout}, P.~A. 1988, \aap, 201,
  311

\bibitem[{{van Zyl} {et~al.}(1981){van Zyl}, {Le}, \& {Amme}}]{vla81}
{van Zyl}, B., {Le}, T.~Q., \& {Amme}, R.~C. 1981, \jcp, 74, 314

\bibitem[{{Vasyunin} {et~al.}(2017){Vasyunin}, {Caselli}, {Dulieu}, \&
  {Jim{\'e}nez-Serra}}]{vc17}
{Vasyunin}, A.~I., {Caselli}, P., {Dulieu}, F., \& {Jim{\'e}nez-Serra}, I.
  2017, \apj, 842, 33

\end{thebibliography}

\end{document}